\documentclass[12pt]{article}
\begin{document}
\title{A Note on Anomalous Effects in 2D  Crystals}
\author{B.G. Sidharth\\
G.P. Birla Observatory and Astronomical Research Centre\\
B.M. Birla Science Centre, Adarsh Nagar, Hyderabad - 500 063
(India)}
\date{}
\maketitle
\begin{abstract}
In this brief note we point out that in the case of two dimensional
nano crystals or quasi-crystals like graphene, there is an
interesting anomalous behaviour of Fermions. This could potentially
be useful. The point is that some electromagnetic properties depend
on the non-commutative geometry of the crystal structure rather than
on the chemistry of the crystal material. Issues related to
magnetism are also commented upon.
\end{abstract}
Recently Andre Geim, who got the joint Nobel Prize for the discovery
of Graphene and his coworkers observed that Graphene is porous only
to protons and opaque otherwise. He hinted that this property could
be used for harvesting hydrogen for use in fuel cells
\cite{bgsijmpa}. On the other hand, the author discussed the magical
and anomalous properties of two dimensional crystals beginning 1995
-- long before the discovery of Graphene
\cite{bgs1,bgs2,bgs3,bgs4,bgs5}. He has also pointed out that these
properties of minimum conductivity and the like are a characteristic
not just of Graphene but of any two dimensional crystal or quasi
crystal, arguing on the following grounds: All this is a consequence
of the noncommutative geometry caused by the lattice like structure
which
is not peculiar to Graphene alone \cite{mac}.\\
The author had gone on to point out that any two dimensional crystal
would exhibit such apparently anomalous properties. Indeed this was
confirmed in the case of Stanene which is a two dimensional crystal
with tin in place
of carbon.\\
Returning to these two dimensional crystals, the protons would be
described by the well known Weyl like two component equation
\cite{greiner}
\begin{equation}
\nu_F \vec {\sigma} \cdot \vec{\nabla} \psi (r) = E \psi
(r)\label{e1}
\end{equation}
Essentially (\ref{e1}) arises as only two of the three momenta $(p_x
p_y p_z)$ would survive owing to the two dimensionality.\\
In the above $\sigma$s are the Pauli matrices and $\nu_F$ is the
Fermi velocity $\sim 10^6m/s$ This is about 300 times less compared
to the velocity of light. However, there is a correspondence between
graphene and the Minkowski world as detailed in an isomorphism
\cite{ijmpe2014} which enables us to go from the one to the other as
far as mathematical formalism is considered. In this case a beam
behaves like a mono energetic beam of Fermions all with energy
$\sim mc^2$ or $m \nu^2_F$ discussed earlier in detail \cite{jsp}.\\
What would happen is that there would be, as discussed earlier, a
Bose Einstein type of condensation of, curiously enough the protons.
To see this in greater detail, we start with the well known formula
for the occupation number of a Fermion gas \cite{huang}
\begin{equation}
\bar n_p = \frac{1}{z^{-1}e^{bE_p}+1}\label{8je9}
\end{equation}
where, $z' \equiv \frac{\lambda^3}{v} \equiv \mu z \approx z$
because, here, as can be easily shown $\mu \approx 1$ (Cf.ref
\cite{jsp,tduniv})
$$v = \frac{V}{N}, \lambda = \sqrt{\frac{2\pi \hbar^2}{m/b}}$$
\begin{equation}
b \equiv \left(\frac{1}{kT}\right), \quad \mbox{and} \quad \sum \bar
n_p = N\label{8je10}
\end{equation}
where the symbols have their usual meaning.\\
Let us consider in particular a collection of Fermions which is
somehow made nearly mono-energetic, as when they stream through a
sheet of graphene or Stanene, that is, given by the distribution,
\begin{equation}
n'_p = \delta (p - p_0)\bar n_p\label{8je11}
\end{equation}
where $\bar n_p$ is given by (\ref{8je9}). (\ref{8je11}) would also
bring us back to the
2D case (Cf. also Appendix).\\
This is not possible in general. By the usual formulation we have,
\begin{equation}
N = \frac{V}{\hbar^3} \int d\vec p n'_p = \frac{V}{\hbar^3} \int
\delta (p - p_0) 4\pi p^2\bar n_p dp = \frac{4\pi V}{\hbar^3} p^2_0
\frac{1}{z^{-1}e^{\theta}+1}\label{8je12}
\end{equation}
where $\theta \equiv bE_{p_0}$.\\
It must be noted that in (\ref{8je12}) there is a loss of dimension
in momentum space, due to the $\delta$ function in (\ref{8je11}).
This is also the case in 2D crystals like
single layer graphene.\\
In an earlier communication \cite{jsp} we showed that in the one
dimensional case, corresponding to nanotubes we would have
\begin{equation}
kT = \frac{3}{5} kT_F\label{Bx}
\end{equation}
where $T_F$ is the Fermi temperature. We can see that for the two
dimensional case too $kT$ would be very small
(Cf.ref.\cite{bgsijmpa}). This is because using the well known
formula for two dimensions we have
\begin{equation}
kT = \frac{e\hbar \pi}{m \nu_F}\label{5}
\end{equation}
\begin{equation}
(kT)^3 = \frac{6e\hbar \nu_F}{\pi}\label{4b}
\end{equation}
Whence we have
\begin{equation}
(kT)^2 = 6 \cdot \nu^2_F \pi^2 m\label{6}
\end{equation}
Remembering that $\nu_F \sim 10^8$, even for a particle whose mass
is that of an electron or a proton, $kT$ in (\ref{6}) is very small.
By way of a comparison for the Fermi temperature we get,
$$kT_F = \frac{\hbar}{2} (z6\pi)^{1/3} \cdot \nu_F$$
We would now have, $kT = <E_p> \approx E_{p}$ because of the mono
energetic feature so that, $\theta \approx 1$.
But we can proceed without giving $\theta$ any specific value.\\
Using the expressions for $v$ and $z$ given in (\ref{8je10}) in
(\ref{8je11}), we get
$$(z^{-1} e^\theta + 1) = (4\pi )^{5/2} \frac{z^{'-1}}{p_0};\mbox{whence}$$
\begin{equation}
z^{'-1}A\equiv z^{'-1}\left(\frac{(4\pi )^{5/2}}{p_0} -
e^\theta\right) = 1,\label{8je14}
\end{equation}
where we use the fact that in (\ref{8je10}), $\mu \approx 1$ as can
be easily deduced.\\
From (\ref{8je14}) if,
$$A \approx 1, i.e.,$$
\begin{equation}
p_0 \approx \frac{(4\pi )^{5/2}}{1+e}\label{8je15}
\end{equation}
where $A$ is given in (\ref{8je14}), then $z' \approx 1$.
Remembering that in (\ref{8je10}), $\lambda$ is of the order of the
de Broglie wave length and $v$ is the average volume occupied per
particle, this means that the gas gets very densely packed for
momenta given by (\ref{8je15}). Infact for a Bose gas, as is well
known, this is the condition for Bose-Einstein condensation at the
level
$p = 0$ (cf.ref.\cite{huang}).\\
In any case there is an anomalous behaviour of the Fermions
accompanied by a "fusion" type effect for protons.\\
On the other hand, if,
$$A \approx 0 (\mbox{that}\quad \mbox{is}\quad \frac{(4\pi )^{5/2}}{e} \approx
p_0)$$
then $z' \approx 0$. That is, the gas becomes dilute, or $V$ increases.\\
More generally, equation (\ref{8je14}) also puts a restriction on
the energy (or momentum), because $z' > 0$, viz.,
$$A > 0(i.e.p_0 < \frac{(4\pi )^{5/2}}{e})$$
$$\mbox{But \quad if}A < 0, (i.e.p_0 > \frac{(4\pi )^{5/2}}{e})$$
then there is an apparent contradiction.\\
The contradiction disappears if we realize that $A \approx 0$, or
\begin{equation}
p_0 = \frac{(4\pi )^{5/2}}{e}\label{8je16}
\end{equation}
(corresponding to a temperature given by $KT = \frac{p^2_0}{2m}$) is
a threshold momentum (phase transition). For momenta greater than
the threshold given by (\ref{8je16}), the collection of Fermions
behaves like Bosons. This is the bosonization effect \cite{schon}.
In this case, the occupation number is given by
$$\bar n_p = \frac{1}{z^{-1}e^{bE_p}-1},$$
instead of (\ref{8je9}), and the right side equation of
(\ref{8je14}) would be given by $' -1'$ instead of $+1$, so that
there would be no contradiction. Thus in this case there is an
anomalous behaviour of the Fermions.\\
The following comment is relevant: It is commonly believed that the
spin features of bosons and fermions are intrinsic properties. That
is true in the usual 3D space, where there is some form of
entanglement with the environment as described e.g. in
ref.\cite{mwt}. Once we deal with the 2D case, this disappears and
we can have ``transmutation" of Fermions
to Bosons or vice versa.\\ \\
\noindent {\large {\bf APPENDIX}(Cf.ref.\cite{bgsfqe})}\\ \\
\noindent To illustrate some of the above statements let us consider
the case of 2D crystals in the context of the fractional Quantum
Hall Effect. We have reiterated that the "graphene" effects are valid for all 2D crystals.\\
The Quantum Hall Effect was discovered experimentally in the 1980s
\cite{klitzing}. In this case, as is by now well known, for a two
dimensional system of electrons, the Hall conductivity is found to
be of the form
\begin{equation}
G = \lambda \cdot \frac{e^2}{h}\label{1}
\end{equation}
where $\lambda$ takes on values
\begin{equation}
\lambda = m/n\label{2}
\end{equation}
 $m$ and $n$ being integers \cite{yenne}.\\
There have been attempts to explain this strange phenomenon from
theory. Particularly by invoking gauge invariance \cite{laughlin}.
 However there have been some persisting puzzles.\\
 We will now look at the Fractional Quantum Hall Effect (FQHE) from a completely novel perspective. Let us consider
 graphene. As is well known this is a single layer graphite with almost magical properties.
 Some of these have been predicted by the author starting 1995 \cite{bgs1995,bgsld,bgsarxiv,jsp}.
 Graphene has a honeycomb like lattice structure so that the space of graphene resembles a
 chessboard with "holes" in space itself as pointed out by Mecklenburg and Regan \cite{mac}.
 This means that there is a fundamental minimum length
 underpinning the system.\\
This fundamental length $L$ leads to a non-commutative geometry as
pointed out by Snyder a long time ago \cite{bgsnap}. This was in the
context of Quantum Electrodynamics \cite{snyder}. This means that if
$(x,y)$ are the coordinates, $xy \ne yx$. A verification for this is
the theoretical deduction of the entire suite of Fractional Quantum
Hall Effect i.e. equation (\ref{C}), as we will see below.\\
Indeed, as pointed out, graphene therefore provides a test bed for
these principles of physics which play a role in Quantum Gravity
approaches. In such a situation it has been shown by the author and
independently Saito \cite{bgsncb,saito,kluwer} that there is a
strong magnetic field. Further, the author showed that this field is
given by
\begin{equation}
B L^2 = hc/e\label{A}
\end{equation}
$L^2$ defines a Quantum of area exactly as in Quantum Gravity
approaches \cite{pstj,bgshj}. This in our case
is the area of individual lattices.\\
To elaborate, the author had argued that (\ref{A}) holds in the case
of a noncommutative geometry. This happens when there is a
fundamental length $L$ which acts as a minimum length of the system.
In this latter case Snyder had shown that commutation relations like
\begin{equation}
[x,y] = \left(\imath L^2/\hbar\right) L_x etc.\label{A1}
\end{equation}
hold good.\\
In these considerations for graphene as is very well known, the
Fermi velocity $\nu_F$ replaces the velocity of light. So we have
for the electron mobility and conductivity
 \begin{equation}
\mu = \nu_F / |E|\label{Aa}
\end{equation}
\begin{equation}
\sigma = (n/A) e \cdot \frac{\nu_F}{|E|}, \, A \sim L^2\label{Ba}
\end{equation}
where $A$, as in the usual theory is the area and $n$ is the number
of electrons. In our case as noted above $A$, the area is made up of
a number of honeycomb lattice areas, each with area $\sim L^2$, that
is
$$A = mL^2$$
where $m$ is an integer.\\
Using these inputs we get (Cf.ref.\cite{bgs3} for details)
\begin{equation}
\sigma = \frac{n}{m} \cdot \frac{e\nu_F}{|B|L^2}\label{B}
\end{equation}
If we now use (\ref{A}) in (\ref{B}) (with $\nu_F$ replacing $c$) we
get for the conductance
\begin{equation}
\sigma = \frac{n}{m} \cdot \frac{e^2}{h}\label{C}
\end{equation}
which defines the fractional Quantum Hall Effect.\\
The author had also shown that it is this non-commutative space
feature in two dimensional structures that explains also Landau
levels \cite{bgsijmpe2005} or the minimum conductivity that exists
in 2D crystals even when there are practically no electrons at the
Dirac points \cite{bgsnap}. In other words several supposedly
diverse phenomena arise from the non-commutative space of these two
dimensional structures.\\
It must be mentioned that the idea of trying to consider graphene
from the perspective of its noncommutative space has been studied by
several authors \cite{bellisard,matilde,menculini,castro}. Some of
the approaches were motivated by the Quantum Electrodynamics in the
spirit of Wilson's Lattice Gauge Theory. We would like to point out
that unlike in the other approaches we have added two new inputs not
used earlier which have lead to the rather comprehensive and neat
deduction of (\ref{C}) like (\ref{A}) and (\ref{A1}). These are
firstly (\ref{A}) which was deduced several years ago in the context
of high
energy physics and quantum gravity and secondly (\ref{x1}), which as shown applies to two dimensional systems.\\
It may be mentioned that over the years the Integral Hall Effect and
Fractional Quantum Hall Effects have not only been observed
experimentally but several excellent simulations exist \cite{exawa}.
Furthermore while it has been known that both the integral and the
fractional effects may be qualitatively related, exact theories to
explain this have not been fully developed. It must be mentioned
that there is another approach that of composite Fermions which
could potentially provide a unified description, for example that of
the approach of J. Jain of Pensylvania
State University. To put it briefly a composite Fermion describes an electron together with an even number of vortices.\\
Finally it may be mentioned that the work of the author and Saito
briefly described above and which shows the production of a magnetic
field due to noncommutative space, provides an explanation for the
Quantum Anomalous Hall Effect which takes place in the absence of an
external magnetic field. This effect was observed recently
\cite{chang}. We can expect that in the process of the fermion boson
transmutation from equation (\ref{8je15})ff., there would be a surge
of magnetism or its sudden disappearance owing to fermion spin
alignments.

\end{document}